\documentclass[10pt,journal,comsoc]{IEEEtran}

\usepackage{graphicx}
\usepackage{cite}
\usepackage{picinpar}
\usepackage{amsmath}
\usepackage{url}
\usepackage{flushend}
\usepackage[latin1]{inputenc}
\usepackage{colortbl}
\usepackage{soul}
\usepackage{multirow}
\usepackage{pifont}
\usepackage{color}
\usepackage{alltt}
\usepackage{hyperref}
\usepackage{enumerate}
\usepackage{epstopdf}
\usepackage{pbox}
\usepackage{enumitem}
\usepackage{fancyhdr}

\usepackage{bm}
\usepackage{amssymb}
\usepackage{stfloats}
\usepackage{algorithmic}
\usepackage[ruled,vlined]{algorithm2e}
\usepackage[table]{xcolor}

\usepackage{wasysym}
\usepackage{diagbox}
\usepackage{tabularx}
\usepackage{caption}
\usepackage{color}
\usepackage{makecell}
\usepackage{booktabs}
\usepackage{multirow}

\usepackage[T1]{fontenc}
\usepackage{graphicx}
\usepackage{booktabs}
\usepackage{array}
\usepackage{enumitem}
\usepackage{hyperref}

\setlist[itemize]{leftmargin=1.1em,topsep=0.2em,itemsep=0.15em,parsep=0em}
\setlist[enumerate]{leftmargin=1.4em,topsep=0.2em,itemsep=0.15em,parsep=0em}

\newcommand{\acceptedDOI}{10.1109/MCOM.001.2600238}
\newcommand{\acceptedDOINote}{\ifx\acceptedDOI\empty DOI: to be added.\else DOI: \acceptedDOI.\fi}

\title{Agentic Peer-to-Peer Networks: From Content Distribution to Capability and Action Sharing}

\author{%
  Taotao~Wang,~\IEEEmembership{Member,~IEEE,}~Lizhao~You,~\IEEEmembership{Member,~IEEE,}~Jingwen~Tong,~\IEEEmembership{Member,~IEEE,}~Chonghe~Zhao,~and~Shengli~Zhang,~\IEEEmembership{Senior~Member,~IEEE}

  \thanks{This is the accepted manuscript of an article accepted for publication in \emph{IEEE Communications Magazine}. \acceptedDOINote}%
  \thanks{T.~Wang,~J.~Tong,~and~S.~Zhang are with the College of Electronics and Information Engineering, Shenzhen University, Shenzhen, China (e-mails: \{ttwang, eejwentong, zsl\}@szu.edu.cn). L.~You is with the Department of Information and Communication Engineering, School of Informatics, Xiamen University, Xiamen, China, and also with the Shenzhen Research Institute of Xiamen University, Shenzhen, China (e-mail: lizhaoyou@xmu.edu.cn). C.~Zhao is with the School of Computer Science and Cyber Engineering, Guangzhou University, Guangzhou, China (e-mail: chonghe.zhao@gzhu.edu.cn). L.~You and J.~Tong are the corresponding authors.}%
}

\begin{document}
\maketitle

\begin{abstract}
The ongoing shift of Artificial Intelligence (AI) from centralized cloud application programming interfaces (APIs) to local inference on edge devices is enabling \textit{Client-Side Autonomous Agents (CSAAs)}---persistent personal agents that can plan, access local context, and invoke tools on behalf of users. As these agents collaborate directly, they form \emph{Agentic Peer-to-Peer (P2P) Networks}. Unlike classic overlays exchanging static, hash-indexed content (e.g., BitTorrent), agentic P2P exchanges dynamic, heterogeneous \emph{capabilities} that are potentially unsafe, requiring semantic discovery, drift management, and risk-aware execution. We outline the networking foundations for such collaboration. To organize this design space, we propose a plane-based reference architecture decoupling connectivity, semantic discovery, and execution. To support discovery under drift, we introduce signed, soft-state capability descriptors to support intent-aware routing. Finally, to cope with adversarial settings, we present a \textit{tiered verification} spectrum---ranging from reputation signals (Tier~1) and canary probes (Tier~2) to cryptographic evidence (Tier~3)---that aligns trust costs with task risks. Results from a proxy-level discrete-event simulation suggest improved workflow success under poisoning and drift with modest control-plane overhead.
\end{abstract}


\section{Introduction}
\IEEEPARstart{T}{he} deployment of {Artificial Intelligence (AI)} is shifting from centralized cloud {application programming interfaces (APIs)} to local execution on edge devices \cite{zheng2024edge_llm_review}, giving rise to \textit{Client-Side Autonomous Agents (CSAAs)}---persistent, personalized systems on smartphones, laptops, and IoT devices that can plan and invoke tools to execute multi-step tasks under user-defined constraints \cite{yao2023react}. 
Yet once agents gain access to files, browsers, and credentials, they become a new security boundary, prompting some organizations to restrict such tooling over concerns about misuse and unintended actions \cite{wired_openclaw_ban_2026}. At the same time, a single CSAA remains bounded by its host's compute budget, tool surface, and local context. To overcome these limits, agents must collaborate---delegating subtasks, exchanging intermediate artifacts, and composing workflows without centralized orchestration \cite{wu2023autogen}.

\begin{figure}[t]
	\centering
	\includegraphics[width=8cm]{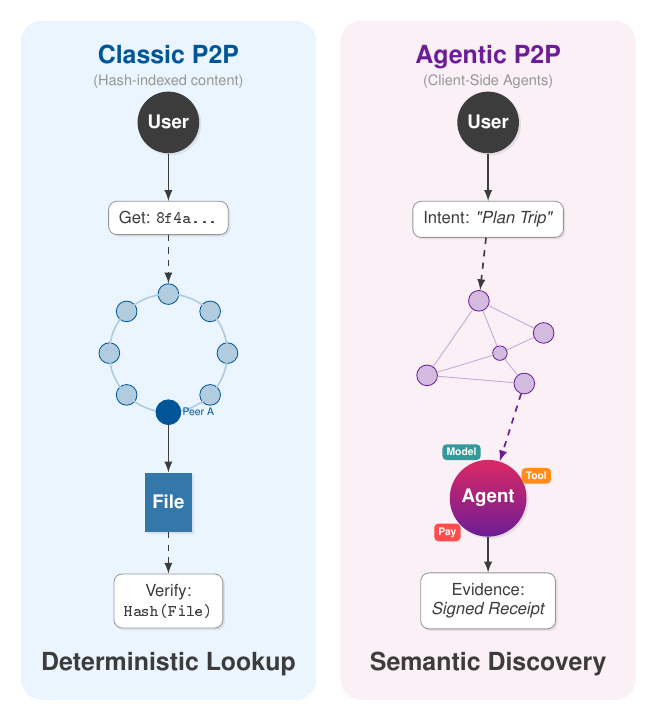}
	\caption{Paradigm shift in P2P networks: classic P2P retrieves static, hash-based content, whereas Agentic P2P discovers dynamic capabilities requiring risk-aware verification.}
	\vspace{-0.4cm}
	\label{fig:paradigm_shift}
\end{figure}

When such collaboration occurs directly between CSAAs to preserve privacy and reduce reliance on centralized platforms, they naturally form \textit{Agentic Peer-to-Peer (P2P) Networks}. We view this setting as a peer-to-peer, application-layer instantiation within the broader Internet of Agents (IoA) vision \cite{wang2025internet_of_agents_survey,ren2026toward_ioa}, which seeks to interconnect heterogeneous agents for dynamic discovery, communication, and task orchestration at scale. {Recent IoA studies have also begun to address capability discovery more explicitly, separating capability announcement from task-oriented discovery \cite{guo2026agent_discovery_ioa}.} At first glance, Agentic P2P resembles classic file-sharing overlays such as BitTorrent \cite{lua2005p2p_survey}. Yet the similarity is only structural. Classic P2P systems distribute \emph{static content}---immutable, hash-addressable, and easy to verify. Agentic P2P instead exchanges \textit{capabilities and actions}: claims such as ``I can generate an image under a strict policy'' or ``I can publish content using a credentialed tool'' are dynamic, heterogeneous, and potentially unsafe to execute. Fig.~\ref{fig:paradigm_shift} visualizes this paradigm shift.

{This positioning separates Agentic P2P from three related lines of work. First, classic P2P overlays and Distributed Hash Tables (DHTs) optimize content lookup and transport over stable identifiers, whereas Agentic P2P must select peers based on semantic capability fit, policy compatibility, freshness, and execution risk. Second, multi-agent systems and orchestration frameworks typically coordinate known agents within managed runtimes or trusted administrative domains \cite{wu2023autogen}; our setting assumes independently operated edge agents in an open overlay, where advertised capabilities may drift or be adversarial. Third, relative to the broader IoA vision \cite{wang2025internet_of_agents_survey,ren2026toward_ioa}, our focus is the P2P networking substrate for decentralized capability discovery and risk-aware collaboration among independently operated agents.}

This shift from content exchange to capability- and action-oriented collaboration necessitates rethinking the application-layer networking substrate. Existing network stacks were designed primarily for data delivery between hosts and processes, rather than for agent collaboration with explicit semantic coordination and verifiable task execution \cite{fleming2025layered_protocol_ioa}. Recent IoA studies likewise organize this problem into layered communication, protocol, management, and coordination functions, reflecting the need to separate connectivity from capability discovery and execution control \cite{ren2026toward_ioa,wang2025internet_of_agents_survey}. Against this backdrop, robust Agentic P2P networking must address three challenges:

\begin{itemize}
\item \textbf{Challenge 1: Semantic discovery beyond exact addressing.}
DHTs excel at exact key lookups \cite{maymounkov2002kademlia} but cannot directly match natural-language intents to suitable executors under constraints \cite{tang2003psearch}. In agentic environments, discovery must operate over semantic relevance, policy compatibility, and execution context rather than immutable file hashes.

\item \textbf{Challenge 2: Verifiable execution beyond data integrity.}
In classic P2P, adversaries mainly degrade availability or authenticity (e.g., via Sybil identities) \cite{douceur2002sybil}. In an execution network, however, malicious peers may misuse tool access, exfiltrate private context, or inject harmful instructions into downstream workflows \cite{liu2023prompt_injection_apps}. Trust must therefore be established through evidence of compliant execution, not merely through object integrity.

\item \textbf{Challenge 3: Capability drift beyond node churn.}
Beyond simple on/off churn, edge agents exhibit \textit{capability drift}: a peer may remain reachable while becoming unable or unauthorized to accept tasks due to battery depletion, policy changes, or evolving local context. The discovery plane must therefore track soft-state availability rather than mere endpoint liveness.
\end{itemize}

We argue that connecting autonomous agents requires a distinct research agenda for the communications and networking community. We make three contributions:

\begin{enumerate}
\item \textbf{A Reference Architecture for Agentic P2P Networking}: Because collaboration couples reachability, discovery, execution, and trust, we present a plane-based architecture that separates connectivity and identity, semantic discovery, and task execution, providing a common networking abstraction for interoperable collaboration across heterogeneous agents.

\item \textbf{A Discovery Model for Dynamic Capabilities}: Because capabilities drift with model version, device state, policy, and context, we introduce signed, soft-state Capability Descriptors (CDs) to represent agent capabilities in a form suitable for intent-aware routing, while Time-to-Live-based refresh naturally captures capability drift in volatile edge environments.

\item \textbf{A Risk-Aligned Trust and Verification Framework}: Because actions differ in privacy exposure and side-effect risk, we replace binary integrity checking with a tiered model of trust, spanning lightweight reputation signals, active canary probes, and strong cryptographic evidence for high-stakes actions.
\end{enumerate}

A proxy-level simulation suggests improved workflow success with near-constant discovery latency and modest control overhead.

The remainder of this article is organized as follows. Sec.~\ref{sec:comparison} contrasts content distribution with capability/action sharing; Sec.~\ref{sec:architecture} introduces the reference architecture; Sec.~\ref{sec:capability_discovery} details capability representation and semantic discovery; Sec.~\ref{sec:trust} presents the verification lifecycle; Sec.~\ref{sec:eval} reports simulation results; and Sec.~\ref{sec:conclusion} discusses open challenges.


\section{From Content Distribution to Action Execution}
\label{sec:comparison}

Classic P2P networks were engineered around objects whose identity, placement, and correctness can be separated from the peer serving them: a file chunk can be named by a hash, replicated opportunistically, and verified after download. By contrast, IoA-style systems shift the unit of exchange toward autonomous agents, semantic communication, and task-driven cooperation \cite{wang2025internet_of_agents_survey}. Agentic P2P sits at this boundary. It preserves the decentralized reachability and self-organization of P2P overlays, but the exchanged object is no longer immutable content; it is a live capability claim whose availability, policy, cost, and safety can change before execution. Table~\ref{tab:comparison} summarizes the resulting shift in network semantics.

\textit{Resource semantics.} In content-centric systems such as BitTorrent and the InterPlanetary File System, resources are static and location-independent. In Agentic P2P, the exchanged object is a stateful capability whose quality may vary with model version, latency, local context, and device state. Discovery therefore requires semantic matching over quickly expiring soft-state descriptors.

\textit{Verifiability and safety.} File integrity is a binary hash check, but agentic execution is probabilistic and may produce side effects. Verification must therefore span reputation, active probing, and strong evidence tiers. Likewise, malicious agents can cause prompt injection, data exfiltration, or delegated-tool misuse. As a result, the network needs least-privilege delegation and sandboxed execution rather than only bandwidth protection \cite{wang2025security_ioa}.

\begin{table}[t]
\caption{Comparison: Classic Content P2P vs. Agentic P2P}
\label{tab:comparison}
\centering
\renewcommand{\arraystretch}{1.05}
\resizebox{\columnwidth}{!}{%
\begin{tabular}{@{}lll@{}}
\toprule
\textbf{Dimension} & \textbf{Classic Content P2P} & \textbf{Agentic P2P} \\ \midrule
\textbf{Core Resource} & Static Files / Chunks & Dynamic Capabilities \& Actions \\
\textbf{Addressing} & Content Hash & Intent \& Constraints (Semantic Matching) \\
\textbf{Verifiability} & Deterministic (Hash Check) & Probabilistic \& Evidence-based \\
\textbf{Cost Model} & Bandwidth Reciprocity & Compute cycles, energy, and service costs \\
\textbf{Failure Mode} & Corrupt Data / Latency & Privacy Leak / Unsafe Action \\
\textbf{Dynamics} & Node Churn (On/Off) & Capability Drift (State/Policy Change) \\ \bottomrule
\end{tabular}%
}
\vspace{-1em}
\end{table}

These constraints make simple ``BitTorrent-for-AI'' indexing insufficient: the network must map ambiguous intents to trustworthy executors, which motivates the reference architecture.

\section{A Reference Architecture for Agentic P2P Networks}
\label{sec:architecture}

\begin{figure}[t]
    \centering
    \includegraphics[width=1\linewidth]{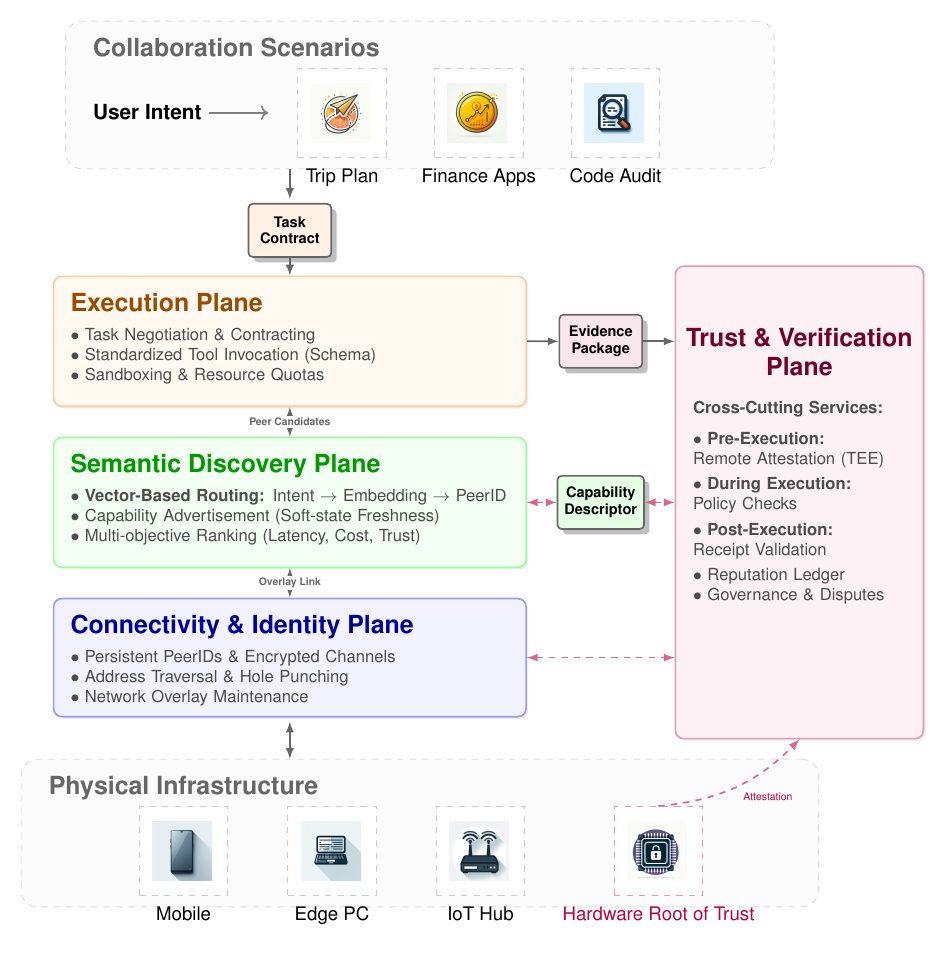}
\caption{Proposed reference architecture. Three horizontal planes (Connectivity, Semantic Discovery, Execution) manage the task lifecycle, while a vertical Trust \& Verification plane enforces security and policy across all layers.}
\vspace{-0.4cm}
    \label{fig:refer_arch}
\end{figure}

While Agentic P2P networks operate at the application layer of the network stack, their internal complexity still calls for a structured decomposition. Recent IoA research has likewise moved toward layered or decomposed views of agent networking, separating communication, protocol, management, coordination, and semantic concerns \cite{wang2025internet_of_agents_survey,ren2026toward_ioa,fleming2025layered_protocol_ioa,xu2026agent_osi}. Motivated by this trend, we propose a \textit{plane-based reference architecture} (illustrated in Fig.~\ref{fig:refer_arch}) that separates connectivity and identity, semantic discovery, execution, and trust/verification. This separation of concerns improves interoperability, allowing heterogeneous agents---differing in model size, toolset, and owner policy---to collaborate within a unified overlay. The architecture consists of four interacting planes.

The planes should be read as a task lifecycle rather than as isolated boxes: a requester first reaches and authenticates peers, then discovers plausible capabilities, then negotiates execution under a risk-appropriate trust policy.

\subsection{Connectivity \& Identity Plane}
This plane abstracts physical-network volatility while providing stable naming and secure channels, so upper layers can reason about peers rather than transient addresses.
\begin{itemize}
    \item \textbf{Persistent Identity}: Agents use cryptographic \textit{PeerIDs} (e.g., public-key hashes) that remain stable as network addresses change; optional signed bindings can link device-local PeerIDs to a user-level alias.
    \item \textbf{Secure Transport}: Given the privacy-sensitive nature of personal-agent tasks, peer-to-peer channels enforce encrypted transport by default.
    \item \textbf{Mobility \& Address-Translation Traversal}: Hole-punching and relaying mechanisms maintain reachable signaling paths as agents move across access networks and private address spaces.
\end{itemize}

\subsection{Semantic Discovery Plane}
This plane maps a natural-language intent to candidate peers. It is where the system moves from ``who is reachable'' to ``who can plausibly perform this task now.''

{In this article, ``agentic routing'' denotes application-layer capability selection rather than packet/path routing. Classical routing still provides reachability once an endpoint is known; the Semantic Discovery Plane maps an intent to peers with compatible capability, policy, resource, and trust posture under freshness and risk constraints.}

\begin{itemize}
    \item \textbf{From Keys to Vectors}: Traditional DHTs support exact key matching; semantic capability discovery instead uses compact descriptors and, optionally, embeddings for fuzzy matching.
    \item \textbf{Similarity-Based Routing}: Queries such as ``analyze a medical PDF'' are routed toward peers whose advertised capability representations best match the request, subject to constraints such as latency, policy, and risk posture.
    \item \textbf{Liveness Management}: To handle capability drift (e.g., devices entering low-power mode or policies changing), this plane uses soft-state advertisements with short Time-to-Live (TTL) values and explicit revocation signals, ensuring the index reflects currently available peers.
\end{itemize}

{Discovery and trust interact in two stages. Public trust hints---reputation summaries, dispute rates, policy tags, and evidence commitments---may filter or rank candidates, but are not proof. After shortlisting, the Execution and Trust planes apply risk-tiered canary probes, attestations, signed receipts, or tool traces; failures trigger fallback and update future reputation or dispute signals.}

\subsection{Trust \& Verification Plane}
This plane manages risk before, during, and after execution. Its role is to make trust evidence proportional to the task rather than forcing every interaction through the same expensive check.
\begin{itemize}
    \item \textbf{Pre-Execution Attestation}: For high-stakes tasks, a requester may require cryptographic evidence (e.g., via {Trusted Execution Environments (TEEs)} such as Intel Software Guard Extensions (SGX) or ARM TrustZone) that the executor runs an expected, policy-constrained kernel.
    \item \textbf{Verifiable Logs}: Signed tool traces and receipts create auditable records without re-executing non-deterministic tasks.
    \item \textbf{Reputation Ledger}: Reputation records reliability signals such as latency, task success, dispute rate, and safety compliance.
\end{itemize}

\subsection{Execution Plane}
The topmost plane handles delegation, messaging, and tool invocation across peers, turning a selected capability into a bounded task with explicit scope and evidence requirements.
\begin{itemize}
    \item \textbf{Task Contracts}: Interactions begin with a negotiated task contract specifying scope, budget (e.g., token or compute limits), deadlines, privacy constraints, and the required verification tier.
    \item \textbf{Standardized Tool Invocation}: A common input/output schema lets agents on different orchestration stacks delegate subtasks through the same execution format.
    \item \textbf{Sandboxing}: Incoming tasks run with capability-based permissions and restricted file/network access, limited to what the contract authorizes.
\end{itemize}

\subsection{Cross-Plane Coordination: Trust-to-Risk Alignment}
A key design principle is \textit{trust-to-risk alignment}: verification rigor should scale with task criticality. Low-risk tasks prioritize responsiveness through lightweight checks, whereas high-risk actions restrict candidate selection and require stronger evidence such as attestation and signed receipts.

{Resource-aware operation is essential in edge P2P settings. A requester should treat compute and energy as first-class constraints rather than afterthoughts: task contracts specify token/compute budgets, deadlines, and verification requirements, while executors accept, reject, or downgrade tasks according to local battery posture, thermal state, accelerator availability, and user policy. These resource signals are intentionally coarse-grained to avoid leaking sensitive device state, but they are sufficient for discovery-time filtering and ranking. Stronger verification also consumes resources; therefore, the tiered verification policy trades additional canary, attestation, and logging cost against task risk.}

\section{Capability Representation and Semantic Discovery}
\label{sec:capability_discovery}

We now instantiate the Semantic Discovery Plane. Given an intent, the network must identify peers satisfying hard constraints (e.g., policy, compatibility) while optimizing soft signals (e.g., relevance, latency, trust). The goal is not to expose every internal detail of an agent, but to publish enough signed, fresh, and policy-aware information for safe shortlisting. We therefore define a standardized capability representation and a baseline discovery workflow, aligned with IoA discussions of capability evaluation, notification, and retrieval \cite{wang2025internet_of_agents_survey}.

\subsection{Signed Capability Descriptors}
The CD is the fundamental discovery unit: a signed, freshness-aware record with a mandatory interoperable core and optional ranking metadata. This keeps the advertisement compact while still leaving room for deployment-specific signals.

\subsubsection{Minimal Core Schema (Mandatory)}
Each valid CD includes:
\begin{itemize}
    \item \textbf{Cryptographic Identity Binding}: \texttt{peer\_id} and a signature over the descriptor.
\item \textbf{Transport Reachability}: Dialable endpoints, plus optional relay-assisted routes for peers behind address translators.
    \item \textbf{Service Definition}: A capability version and schema-hash commitment.
    \item \textbf{Policy Posture}: Machine-readable tags (e.g., \texttt{no-logs}, \texttt{sandbox-only}) for pre-connection filtering.
    \item \textbf{Liveness Metadata}: A creation timestamp and a short TTL value to enforce soft-state consistency and mitigate capability drift.
\end{itemize}

{
A compact formalization is
\texttt{CD := <peer\_id, sig, endpoints, cap\_id, schema\_hash, policy\_tags, ts, ttl, metadata>},
where the first eight fields constitute the mandatory interoperable core and
\texttt{metadata} carries optional semantic, quality-of-service (QoS), and trust hints. A descriptor is
considered admissible only if signature verification succeeds, the advertised
schema hash matches the claimed capability version, and the TTL has not expired.
}

\subsubsection{Semantic and Performance Extensions (Optional)}
Optional metadata provides heuristic ranking signals, subject to later verification:
\begin{itemize}
    \item \textbf{Semantic Signals}: Embeddings or keyword tags describing domain expertise.
    \item \textbf{Performance Hints}: {Self-declared metrics such as expected latency class, compute/accelerator availability, battery or charging posture, thermal/load state, and coarse task-cost estimates.}
    \item \textbf{Evidence Commitments}: Hashes or pointers to attestations and audit logs for pre-validation.
\end{itemize}

{Because CDs are public, discovery exposes only coarse capability tags, protocol compatibility, policy posture, freshness, and commitments to richer schemas or evidence. Fine-grained tool schemas, credential state, detailed telemetry, private prompts, and audit artifacts stay local until an encrypted handshake and task contract define purpose, scope, and verification tier \cite{wang2025internet_of_agents_survey}.}


\subsection{Protocol Sketch: Registry-Based Discovery}
As a baseline instantiation, a verifiable community registry uses soft-state consistency. The registry is not essential to the architecture, but it makes the control flow explicit:

{Architecture-level invariants are signed CDs, freshness and revocation semantics, constraint-aware selection, task contracts, and tiered verification; none requires a logically centralized registry. Registry-specific pieces are only the \texttt{PUBLISH}/\texttt{REFRESH}/\texttt{REVOKE} API, indexing/lookup service, and simulator latency/overhead parameters.}

\begin{enumerate}
    \item \textbf{Publication}: A provider sends signed \texttt{PUBLISH(CD, TTL)}; the registry validates and indexes it until expiry.
    \item \textbf{State Maintenance}: Periodic \texttt{REFRESH} updates let drifted capabilities naturally expire.
    \item \textbf{Revocation}: \texttt{REVOKE} removes withdrawn or compromised capabilities.
    \item \textbf{Intent-Based Querying}: A requester submits an intent vector plus hard constraints.
    \item \textbf{Filtering and Ranking}: The engine removes constraint violations and ranks candidates by semantic, performance, and trust signals.
\end{enumerate}

{
Operationally, this workflow follows a three-stage selection pipeline:
\emph{filter}, \emph{score}, and \emph{shortlist}. The filter stage removes CDs
that fail signature, freshness, policy, or protocol-compatibility checks. The
score stage then assigns each remaining peer a requester-defined utility, e.g.,
$U(p \mid q) = \lambda_s \mathrm{Sim}(q,p) + \lambda_q \mathrm{QoS}(p) +
\lambda_t \mathrm{TrustHint}(p) - \lambda_c \mathrm{Cost}(p)$, where hard
constraints are enforced before scoring rather than traded off. Here
$\mathrm{Cost}(p)$ may include compute load, expected token or energy use,
monetary/tool cost, and verification overhead. Finally, the top-$K$ candidates
are returned as a shortlist for verification-time fallback in the Execution
Plane.
}

The output is a Pareto set passed to the Execution Plane for risk-aware verification (Section~\ref{sec:trust}).

\subsection{Implementation Considerations}
The architecture is substrate-agnostic: permissionless networks may use DHT indexing, while enterprise clusters may use federated registries. In either case, robustness depends on soft-state TTLs to avoid ``zombie'' peers and on admission controls plus verification feedback to resist index poisoning.

Ultimately, the Semantic Discovery Plane produces a set of candidates that are claimed to be capable but remain untrusted by default. This gap between discoverability and validated semantic correctness echoes recent arguments that agent protocols must move beyond well-formed message exchange toward explicit semantic validation before execution begins \cite{fleming2025layered_protocol_ioa}.

\section{Trust Establishment and Verifiable Execution}
\label{sec:trust}

Semantic discovery identifies candidates but not their trustworthiness. In open overlays, peers may overclaim capabilities, free-ride, or misuse delegated tools, so trust must be established dynamically before sensitive context is shared and audited after execution.

{
\textit{Threat model and countermeasure mapping.} We assume an open overlay in which adversarial peers may create Sybil identities, publish inflated or stale capability descriptors, overclaim task competence, return low-effort results, inject malicious instructions, attempt data exfiltration, misuse delegated tools, or collude to manipulate reputation and trust scores. We do not assume that public capability claims are truthful, but we assume standard cryptographic primitives are not broken and that signed evidence or attestation artifacts cannot be forged without key or TEE compromise. The countermeasures are mapped to the task lifecycle: signed CDs, TTL refresh, and revocation limit descriptor tampering and staleness; admission controls and verification feedback reduce index poisoning; Tier~2 canary probes and fallback selection expose capability overclaiming; task contracts, least-privilege sandboxing, and restricted tool access reduce exfiltration and delegated-action abuse; and signed evidence packages, dispute markers, and multi-dimensional reputation updates make collusion and trust-system manipulation more costly and auditable.
}

\subsection{Tiered Verification Framework}
To balance overhead and assurance, verification uses three policy-driven tiers negotiated in the task contract.
\begin{itemize}
    \item \textbf{Reputation-Based for Low Risk Task (Tier 1)}: Public or stateless tasks rely on local history or a federated ledger, with optional signed receipts for reputation updates.
    \item \textbf{Interactive Challenge-Response for Medium Risk Tasks (Tier 2)}: Moderate-risk tasks require lightweight puzzles or ``canary'' microtasks before full execution, filtering lazy or poisoned candidates with limited latency.

    \item \textbf{Evidence-Based Execution for High Risk Task (Tier 3)}: High-stakes tasks involving private inputs or irreversible effects require strong evidence such as remote attestation and auditable tool traces \cite{xu2026agent_osi}.
\end{itemize}


\subsection{Protocol Workflow: The Verification Lifecycle}
The end-to-end workflow negotiates evidence upfront and validates it after execution.

\begin{enumerate}
    \item \textbf{Contract Negotiation}: A Task Contract specifies the \texttt{task\_id}, verification tier, resource budgets, and policy constraints; the executor accepts only if local policy permits.
    
    \item \textbf{Pre-Execution Validation}: 
    The requester issues canary microtasks for Tier~2 or validates an attestation quote for Tier~3 before transmitting sensitive payloads.
    
    \item \textbf{Sandboxed Execution}: The executor runs the task in WebAssembly or a microVM, with external API calls intercepted and logged.
    
    \item \textbf{Evidence Submission}: The executor returns the result with a signed Evidence Package:
    \begin{itemize}
        \item \textit{Receipts}: timestamps and signatures over result metadata.
        \item \textit{Tool Traces}: ordered, hashed tool logs with policy-based redaction.
        \item \textit{Attestation (Tier 3)}: binding the result to the secure enclave.
    \end{itemize}
    
    \item \textbf{Settlement and Feedback}: Valid evidence updates trust; missing or inconsistent evidence triggers dispute markers.
\end{enumerate}

\begin{figure}[t]
    \centering
    \includegraphics[width=0.75\linewidth]{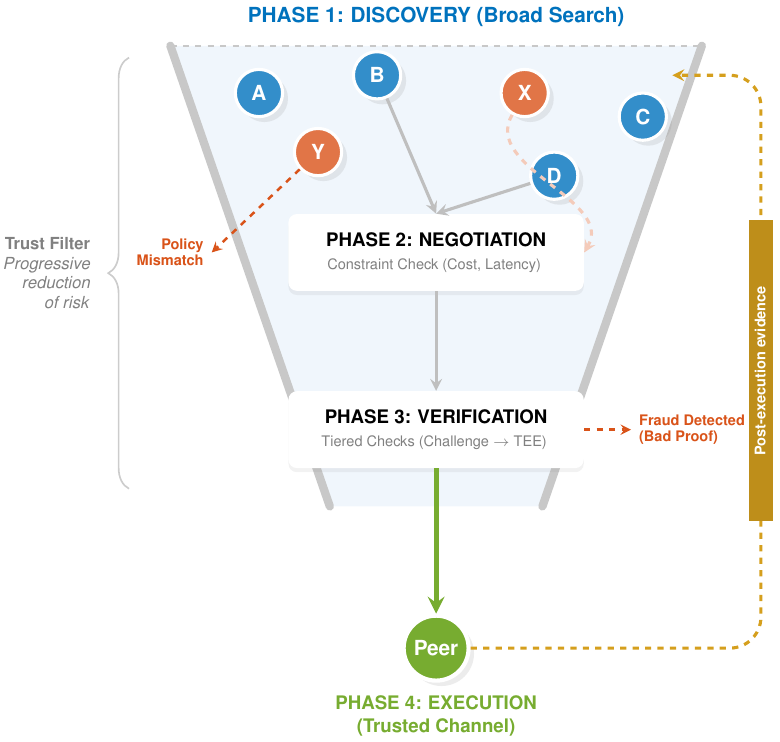}
\caption{Trust establishment funnel: candidates are narrowed through negotiation and tiered verification before execution.}
    \label{fig:trust_funnel}
    \vspace{-0.4cm}
\end{figure}

Fig.~\ref{fig:trust_funnel} shows how candidates are narrowed through constraint checks and tiered verification so overhead remains proportional to risk.

\section{Simulation Investigations}
\label{sec:eval}

To provide a preliminary feasibility check of the proposed architecture, we implemented the protocol sketches from Sections~\ref{sec:capability_discovery}--\ref{sec:trust} in a discrete-event simulator (source code: \url{https://github.com/plan-lab-szu/agentic_p2p_sim}). The simulator captures soft-state CDs with TTL-based expiry/refresh, registry-based discovery, Sybil-style index poisoning (misleading CDs), and risk-aligned verification with Tier~2 canary probes and fallback selection. {This simulator should be treated as an architecture-level sensitivity study rather than a benchmark of any specific model stack, discovery engine, or deployment environment. The public repository includes a model README, dependency file, JSON configurations, seeded Monte Carlo entry points, scripts for the parameter sweeps used to generate the reported simulation results, and CSV/JSON summaries for regenerating the reported curves.}

\subsection{Workflow and Baselines}
We evaluate \textit{Decentralized Creative Publishing}: ``Generate a marketing poster and publish it.'' The workflow is intentionally small, but it combines two stresses that appear in real agent collaboration: compute-heavy generation and a credentialed publishing action with side effects. It therefore requires an \textit{Image Generator} and a credentialed \textit{Publisher}, stressing discovery under poisoning and verification before irreversible actions.

{We use \textit{No Verify} as the baseline comparator: it directly selects the top discovery candidate without preflight probing. The proposed \textit{Risk-Aware} strategy instead applies Tier~2 canary challenge-response to the top candidates for the generator step with fallback on failures, and requires an evidence package (signed receipts/traces) for the publisher step.}

\subsection{Simulation Setup}
Each run captures background CD maintenance and concurrent workflows, averaged over independent seeds. Fuzzy intent--capability matching is a scalar semantic score advertised in CDs and inflated by Sybil poisoning to manipulate the Top-$K$ shortlist. {This scalar score is a model-agnostic proxy for intent--capability affinity rather than the output of any particular embedding pipeline.} It abstracts away embedding-model/version sensitivity, prompt and context dependence, semantic false positives/negatives, and retrieval-pipeline effects such as indexing, reranking, calibration, and privacy filtering. We instantiate $N \in [20,200]$ CSAAs and vary the Sybil fraction $\alpha$. Delays follow a heterogeneous Wi-Fi/cellular model, registry lookup adds lightweight $O(\log_2 N)$ processing, CDs use TTL refresh, and drift appears as state-version changes that increase stale-descriptor failure. Control-plane overhead excludes application payload.

{
\textit{Scope and limitations.} The simulator is an architecture-level sensitivity study, not a packet-, power-, or model-execution benchmark. Drift is modeled as stochastic state-version changes: a reachable peer's model, tool permissions, local context, battery/thermal posture, or policy may no longer match its last CD. This isolates how TTL refresh, canary probing, and fallback react to stale or misleading advertisements; it does not model quality drift, operating-system scheduling, accelerator contention, or thermal throttling. In deployment, these constraints enter through coarse Performance Hints, task-contract budgets, and the $\mathrm{Cost}(p)$ term, with detailed telemetry kept local or revealed only after a secure handshake. Trace-driven power/thermal measurements are left as future work.
}

\begin{figure*}[t]
    \centering
    \includegraphics[width=1\linewidth]{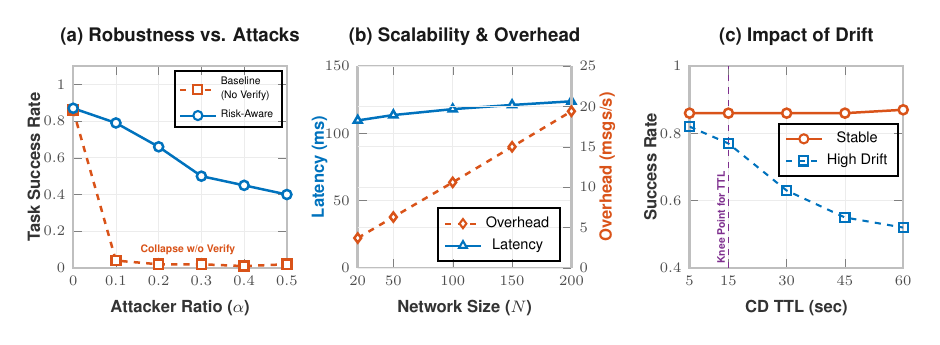}
\caption{Performance of the decentralized creative publishing workflow. (a) Success rate vs.\ Sybil ratio $\alpha$. (b) Discovery latency and control-plane overhead vs.\ network size $N$. (c) TTL sensitivity under capability drift.}

\vspace{-0.4cm}
    \label{fig:performance}
\end{figure*}

\subsection{Results and Discussion}

\subsubsection{Robustness under Sybil poisoning}
Fig.~\ref{fig:performance}(a) sweeps attacker ratio $\alpha$ from 0 to 0.5 ($N{=}100$, $\mathrm{TTL}{=}15$\,s, drift disabled). \textit{No Verify} drops as poisoned CDs dominate the shortlist. {At $\alpha{=}0.2$, the baseline falls to roughly 0.02 end-to-end workflow success, whereas \textit{Risk-Aware} retains about 0.66; even at $\alpha{=}0.5$, \textit{Risk-Aware} still achieves about 0.40.} Tier~2 canaries and fallback filter deceptive peers before expensive execution.
{This confirms that semantic relevance alone is insufficient in open overlays; ranking needs lightweight probes before costly or irreversible execution.}

\subsubsection{Scalability and control-plane overhead}
Fig.~\ref{fig:performance}(b) sweeps $N$ from 20 to 200 (fixed $\alpha{=}0.2$, $\mathrm{TTL}{=}15$\,s, drift disabled). {Across this range, median discovery latency increases only from about 109.6\,ms to 123.8\,ms, while control-plane overhead rises from about 3.7 to 19.4 messages/s.} Thus, registry processing adds modest delay and discovery maintenance remains manageable for moderate swarms.
{The weak latency growth comes from verifying only a bounded shortlist; overhead mainly reflects refresh and workflow-control traffic.}

\subsubsection{TTL sensitivity under capability drift}
Fig.~\ref{fig:performance}(c) sweeps $\mathrm{TTL}\in\{5,15,30,45,60\}$\,s. {In the stable regime, workflow success stays near 0.86 across all TTL values; under \textit{High Drift}, it declines from about 0.82 at 5\,s to about 0.52 at 60\,s.} Short TTLs improve freshness at higher refresh cost; $\mathrm{TTL}{\approx}15$\,s balances both.
{Thus TTL is a policy knob: faster drift calls for shorter refresh intervals, while stable clusters can trade freshness for lower control traffic.}

\section{Conclusion and Future Challenges}
\label{sec:conclusion}

Agentic P2P networks transform decentralized overlays from content distribution into capability and action sharing, where failures can cause privacy or safety harm rather than only wasted bandwidth. We proposed a plane-based architecture with signed soft-state descriptors, semantic discovery, and tiered verification. Simulations under poisoning and drift show that adaptive verification improves workflow reliability while keeping discovery latency and control overhead modest. Open challenges remain:
interoperable descriptor, contract, and evidence schemas; privacy-preserving discovery through coarse tags, commitments, and on-demand disclosure; and incentive and accountability mechanisms for edge compute, energy use, and disputes.
Together, these directions outline a path toward privacy-preserving and robust Internet-of-Agents services on P2P substrates.

\bibliographystyle{IEEEtran}
\bibliography{ref}

\end{document}